\documentclass[twocolumn,amsmath,amssymb]{revtex4}

\pdfoutput=1
\usepackage{graphicx}
\usepackage{txfonts}

\begin{document}

\title{The scaling properties of dynamical fluctuations in temporal networks}

\author{Liping Chi}
\email{chilp@mail.ccnu.edu.cn} \affiliation{College of Physical Science and Technology, Central China Normal University, Wuhan
430079, China}
\author{Chunbin Yang}
\affiliation{College of Physical Science and Technology, Central China Normal University, Wuhan
430079, China}

\date{August 8, 2015}

\begin{abstract}
The factorial moments analyses are performed to study the scaling properties of the dynamical fluctuations of contacts and nodes in temporal networks based on empirical data sets. The intermittent behaviors are observed in the fluctuations for all orders of the moments. It indicates that the interaction has self-similarity structure in time interval and the fluctuations are not purely random but dynamical and correlated. The scaling exponents for contacts in {\bf\it Prostitution} data and nodes in {\bf\it Conference} data are very close to that for 2D Ising model undergoing a second-order phase transition.
\end{abstract}

\maketitle

\section{Introduction}

Interactions in complex systems are not static but change over time, which can be modelled in terms of temporal networks \cite{Holme2012}. Temporal network consists of a set of contacts $(n_i, n_j, t)$, emphasizing on the time {\it when} node $i$ and $j$ have a connection. The addition of time dimension provides a new sight into the framework of complex network theory. In temporal networks, both structural properties and spreading dynamics crucially depend on the time-ordering of links.

The research of temporal networks has attracted great attention and it mainly focuses on two major aspects from the point view of time dimension. One is corresponding to the strategy of time aggregation especially when the topological characteristics are more relevant than the temporal properties. The topological structure of temporal network is achieved through aggregating contacts over a certain time interval and the temporal network is then represented as a series of snapshots of static graphs. Consequently, many existing concepts and tools of static graphs can be adopted to analyze temporal networks, since it is usually easier to analyze static networks. For example, the degree of a node $k_i(t)$ is described as the number of links that it has to other nodes within the time window $[t, t+\Delta t]$. The error and attack strategies in static networks have been applied to evaluate the temporal vulnerability \cite{Trajanovski}, and so on. In order to understand the structure of temporal networks, it plays a crucial role to choose an optimal time interval $\Delta t$ to construct static graphs from temporal networks. Krings {\it et al.} studied the influences of time intervals when aggregating the mobile phone network over time \cite{Krings}. Holme analyzed three ways of constructing static snapshots from temporal networks \cite{Holme2013}, but no candidate weighing out as a best choice. It is now still an open question on how to choose the time interval to represent temporal networks.

The other aspect is related to using dynamical processes to probe into the influence of time series on temporal network. We should take into account the time-ordering of each contact and the inter-event time between two consecutive contacts. The inter-event time distribution in temporal network follows a power-law, which is also called burstiness \cite{Barabasi2005}. Although it is recognized that time-ordering and bursty characters have strong influences on the dynamical processes of temporal networks, numerous studies have appeared to arrive at contradictory results. Lambiotte {\it et al.} had stressed that time-ordering and burstiness of contacts were critical in spreading process, which leaded to slow down spreading \cite{Lambiotte}. In the work of Rocha {\it et al}, they concluded that temporal correlations accelerated outbreaks \cite{Rocha2011} in SI and SIR model. Miritello {\it et al.} demonstrated that bursts hindered propagation at large scales, but group conversations favored local rapid cascades \cite{Miritello}.

Despite the promoting results in temporal networks, this field is still in its early stages about how temporal effect and topological structure interplay and hence affect the dynamical process. In this paper, based on empirical data sets, we will investigate the scaling properties of the dynamical fluctuations of contacts and nodes in temporal networks by using the factorial moments. We are aiming at extracting the fundamental properties from the large amount of data and revealing the influences of time effects on temporal networks from a new perspective.

The rest of the paper is organized as follows. Section II briefly introduces the method of factorial moments. In Section III we give a brief description of the data sets and present the corresponding results, especially the scaling properties of fluctuations for contacts and nodes in the empirical data sets. Conclusions are offered in the final section.

\section{Method of factorial moments}

Temporal network consists of a sequence of contacts $(n_i,n_j,t)$, representing that node $i$ and node $j$ has a contact at time $t$. The number of contacts characterizes the frequency that individuals are connected with each other and the number of nodes describe the activeness that individuals are involved. In this paper, factorial moments will be used to study the dynamical fluctuations of contacts and nodes in temporal networks and the scaling properties of those fluctuations in the system.

Factorial moments are originally introduced in nuclear physics to study the multiplicity fluctuation of hadrons produced during the high energy collisions\cite{Bialas1986}. The fluctuations and correlations in multiplicity distributions provide a general and sensitive method to characterize the dynamical interactions. Here we will focus on the multiplicity of contacts and nodes in temporal networks. Consider the time series of contacts (or nodes) $y(t)$, where $t$ is the time that contacts happen and $t$ ranges from 0 to $T$. We divide the whole time range $T$ into $M$ equal bins (the remainders are discarded). So the time interval in each bin is $\Delta t = T/M$. Within each bin window $m$ ($m=1,2,...,M$), denote the number of contacts (or nodes) as $n_m$. Of course, $n_m$ fluctuates for different bin windows. To measure the fluctuations and correlations, the $q-th$ order factorial moment is introduced as,
\begin{align}
f_q & = \frac{1}{M}\sum\limits_{m=1}^{M}n_m(n_m-1)...(n_m-q+1) \notag \\
& = \langle n_m(n_m-1)...(n_m-q+1) \rangle.
\label{factmoment}
\end{align}

In factorial moments, $f_1=\langle n \rangle$ is the mean number of contacts (or nodes) under a certain bin size, averaged over all the bins $m$. Note that $n_m$ must be greater than $q$ $(n_m > q)$ in order to contribute to $f_q$ , and $q$ is usually an integer. As $M$ increases, $\Delta t$ is decreased and the average multiplicity $\langle n \rangle$ in a bin decreases. This may lead to $ n_m < q$ which is not allowed. Thus high $q$ corresponds to higher $n_m$ in the bin under consideration, i.e., large fluctuations from $\langle n \rangle$ \cite{Hwa1998}.

Normalized factorial moments are more generally used,
\begin{equation}
F_q=\frac{f_q}{f_1^q}.
\end{equation}

\noindent It can be proved that $F_q$ can filter out the statistical fluctuations. The method of factorial moments has been applied to analyze different complex systems, such as multiplicity of produced hadrons \cite{Chunbin1998}, human electroencephalogram and gait series in biology \cite{Hwa2002,Yang2002}, financial price series \cite{Schoeffel}, critical fluctuations in Bak-Sneppen model \cite{Xiao}, spectra analysis of complex networks \cite{Yang2005}, to name a few. Specially it indicates that the fluctuations in the system have self-similarity when $F_q$ has a power-law dependence on the bin size $M$.
\begin{equation}
F_q \propto M^{\alpha_q}, \qquad \alpha_q>0.
\end{equation}

\noindent This phenomenon is referred to as the intermittency. Intermittency basically means random deviations from smooth or regular behavior. Intermittent behavior is expected in a variety of statistical systems at the phase transition point of the second-order type. Hence the existence of intermittency suggests that the fluctuations are not purely Poisson distribution, but the indication of dynamical processes in the fluctuations.

\section{Results and discussions}

In this paper the factorial moments analyses are performed to uncover the scaling properties of the fluctuations in temporal networks based on the following two empirical data sets.

\begin{itemize}
  \item[] {\bf\it Prostitution:} The data set consists of sexual contacts between sex buyers and sellers from a Brazilian web forum \cite{Rocha2010}. The time resolution is 1 day and the whole time range is $T = 2232 $ days.

  \item[] {\bf\it Conference:} The data set was collected at a 3-day  conference from face-to-face interactions between conference participants. A contact is recorded every 20-second intervals if two individuals are within range of 1.5m \cite{Isella}. The whole time range is $T= 212 340$ seconds.

\end{itemize}

We now divide the whole time range $T$ into $M$ bins and count the number of contacts and nodes in each bin window. Calculate $f_q$ and $F_q$ according to Eq. (1) and (2), respectively. It is noticed that $f_q$ is averaged over all bins (known as the horizontal average).

Figure 1 presents the log-log plot of $F_q$ as a function of $M$ for contacts (open circles) and nodes (filled circles) in {\bf\it Prostitution} data. With $M$ ranging from about 3 to 60 bins, it means that the time interval $\Delta t$ extends approximately from 30 to 750 days. We find that $\ln F_q$ increases linearly with $\ln M$ for both contacts and nodes by $q$ varying from 2 to 6. The slopes of nodes are a little larger than that of contacts. The same phenomena have also been observed in {\it Conference} data in Fig. 2.

The increase of bin size $M$ means that the fluctuations of arbitrary sizes can appear in the system, and consequently leading to the growth of factorial moment $F_q$ with $M$. The scaling relationship between $F_q$ and $M$, $F_q \sim M^{\alpha_q}$, indicates the existence of intermittency. As stated in Ref. ~\cite{Schoeffel}, for uncorrelated Poissonian or Gaussian distributions, $F_q=1$ for all orders $q$; whereas for correlated contacts or nodes distributions, $F_q$ should increase with the growth of bin size $M$. Hence the intermittent behavior implies that the fluctuations of contacts and nodes in both {\it Prostitution} and {\it Conference} data have self-similar structures and the fluctuations are not random Poisson distribution but have dynamical and correlated behaviors inside.

\begin{figure}
     \includegraphics[width=0.45\textwidth]{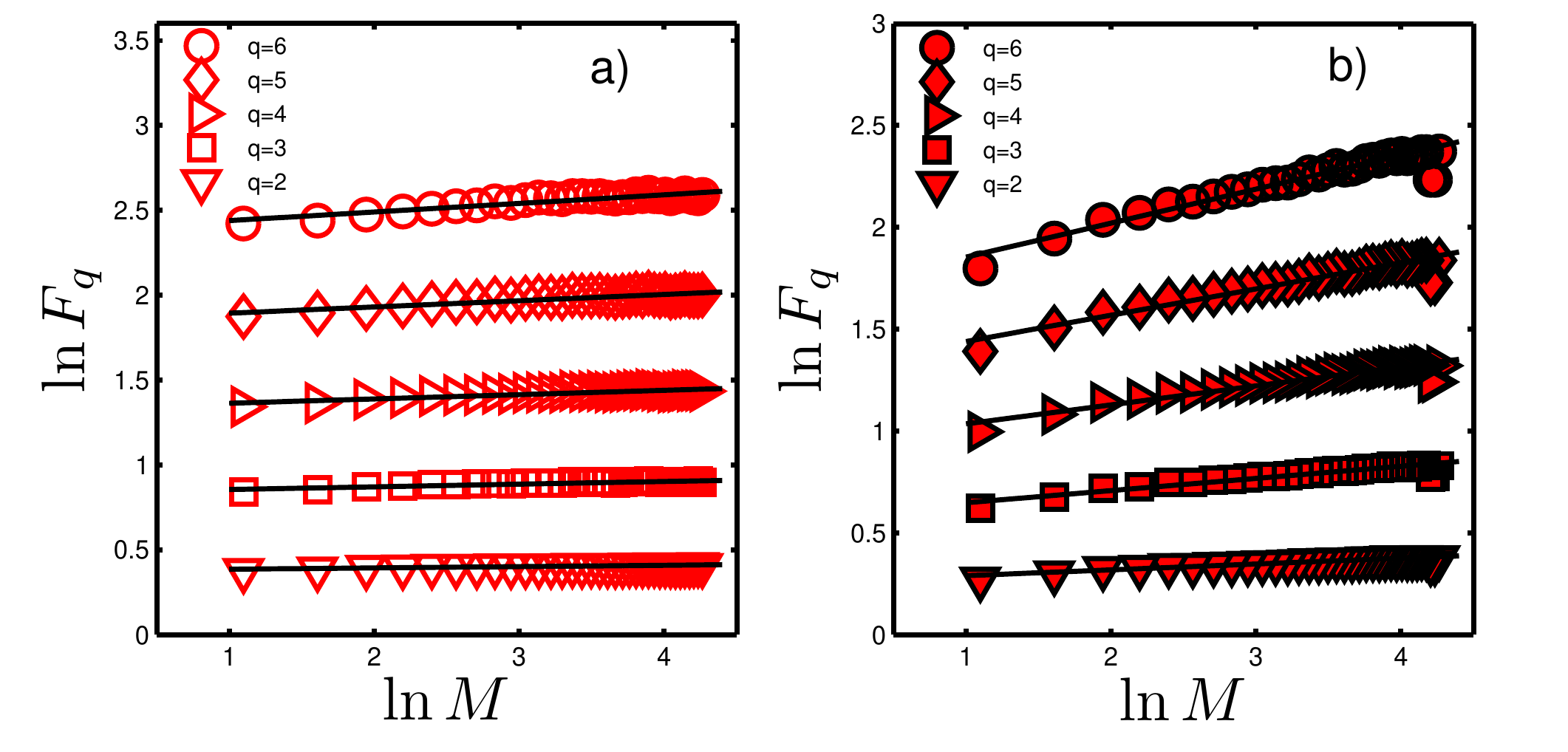}
     \caption{Log-log plot of factorial moments $F_q$ as a function of bin size $M$ for {\bf\it Prostitution} data with $q$ varying from 2 to 6. (a) The fluctuations of contacts (open circles). (b) The fluctuations of nodes (filled circles). The symbols are the same below.}
\end{figure}

\begin{figure}
     \includegraphics[width=0.45\textwidth]{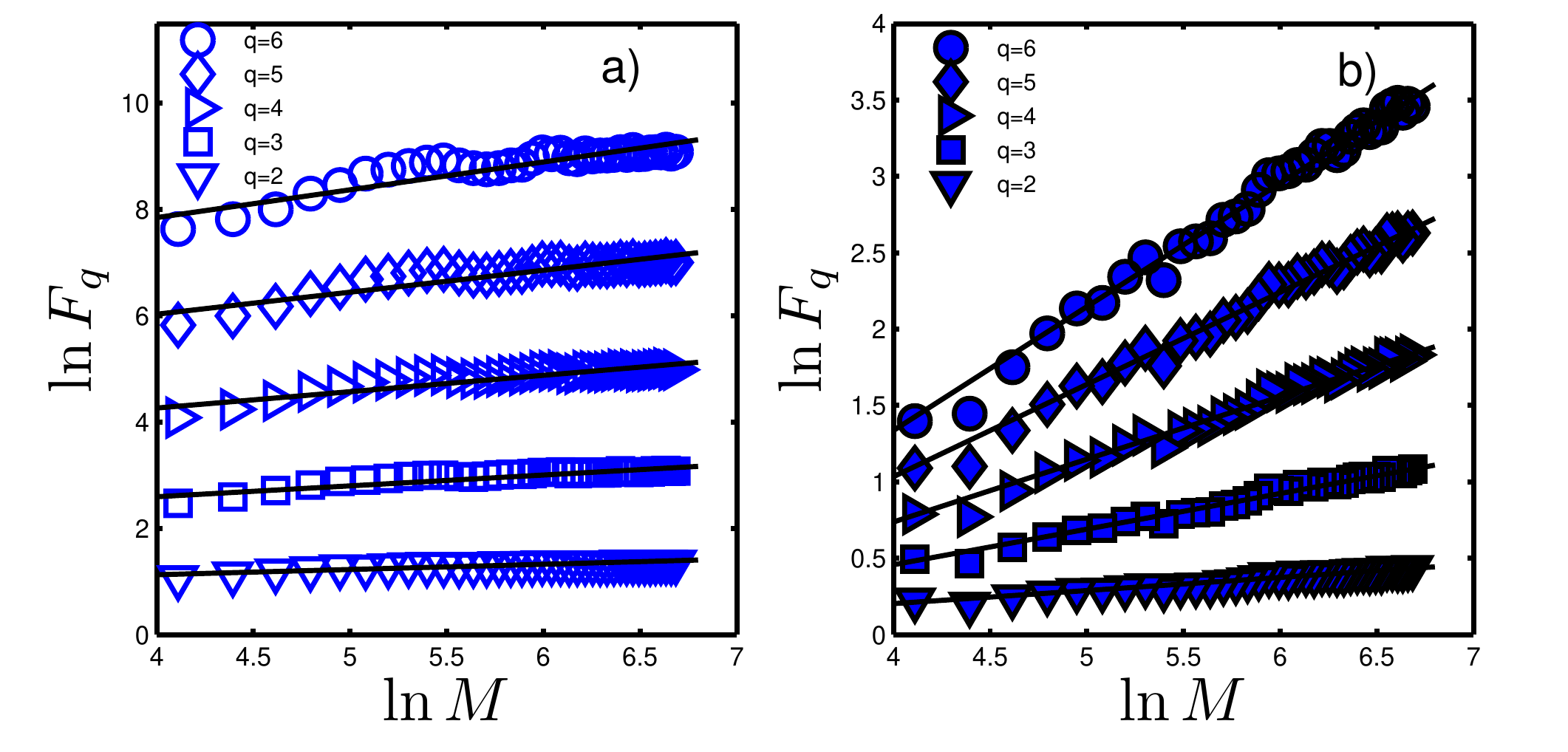}
     \caption{Log-log plot of $F_q$ as a function of $M$ for {\bf\it Conference} data with the range of $q$ from 2 to 6. (a) for contacts; (b) for nodes.}
\end{figure}

Further investigations have been performed on $F_q$ and $F_2$. The scaling between $F_q$ and $F_2$ is more general than intermittency, which could be true even under the condition that intermittency does not exist.

We plot $F_q$ as a function of $F_2$ on the log-log scale for {\it Prostitution} and {\it Conference} data sets in Fig. 3 and 4, respectively. The scaling relationship between $F_q$ and $F_2$ can be clearly observed in both figures.
\begin{equation}
F_q \propto F_2^{\beta_q},
\end{equation}

\noindent where $\beta_q = \alpha_q/\alpha_2$ for the case of intermittency. We are interested more in the dependence of $\beta_q$ on $q$. The plot of $\beta_q$ as a function of $(q-1)$ is presented on a log-log scale in Fig. 5 for {\it Prostitution} data and in Fig. 6 for {\it Conference} data. There is a remarkably linear relationship between $\beta_q$ and $(q-1)$ for all $q$. Now one has
\begin{equation}
\beta_q \propto (q-1)^{\gamma}.
\end{equation}

The linear fits are also plotted in the figures.  In {\it Prostitution} data, $\gamma$ are 1.341 for contacts and 1.104 for nodes. In {\it Conference} data $\gamma$ are 0.992 and 1.345 for contacts and nodes, respectively. It should be recognized that the power-law relationship in Eq. (5) implies that the exponents $\beta_q$ are independent of bin size $M$. It suggests a common feature of scaling invariance in temporal networks.

\begin{figure}
    \centering
     \includegraphics[width=0.45\textwidth]{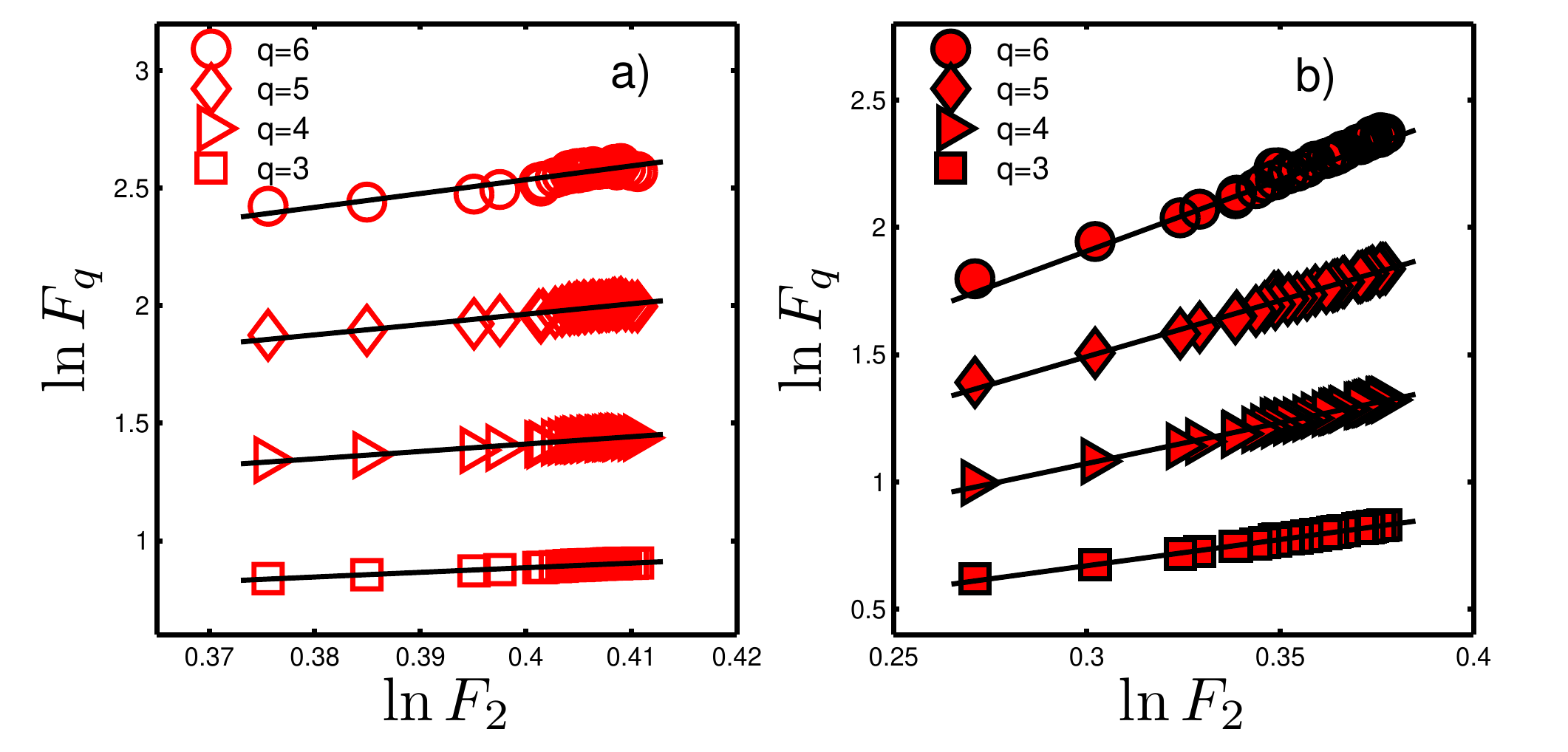}
     \caption{Log-log plot of $F_q$ as a function of $F_2$ for {\bf\it Prostitution} data. (a) for contacts; and (b) for nodes.}
\end{figure}

\begin{figure}
    \centering
     \includegraphics[width=0.45\textwidth]{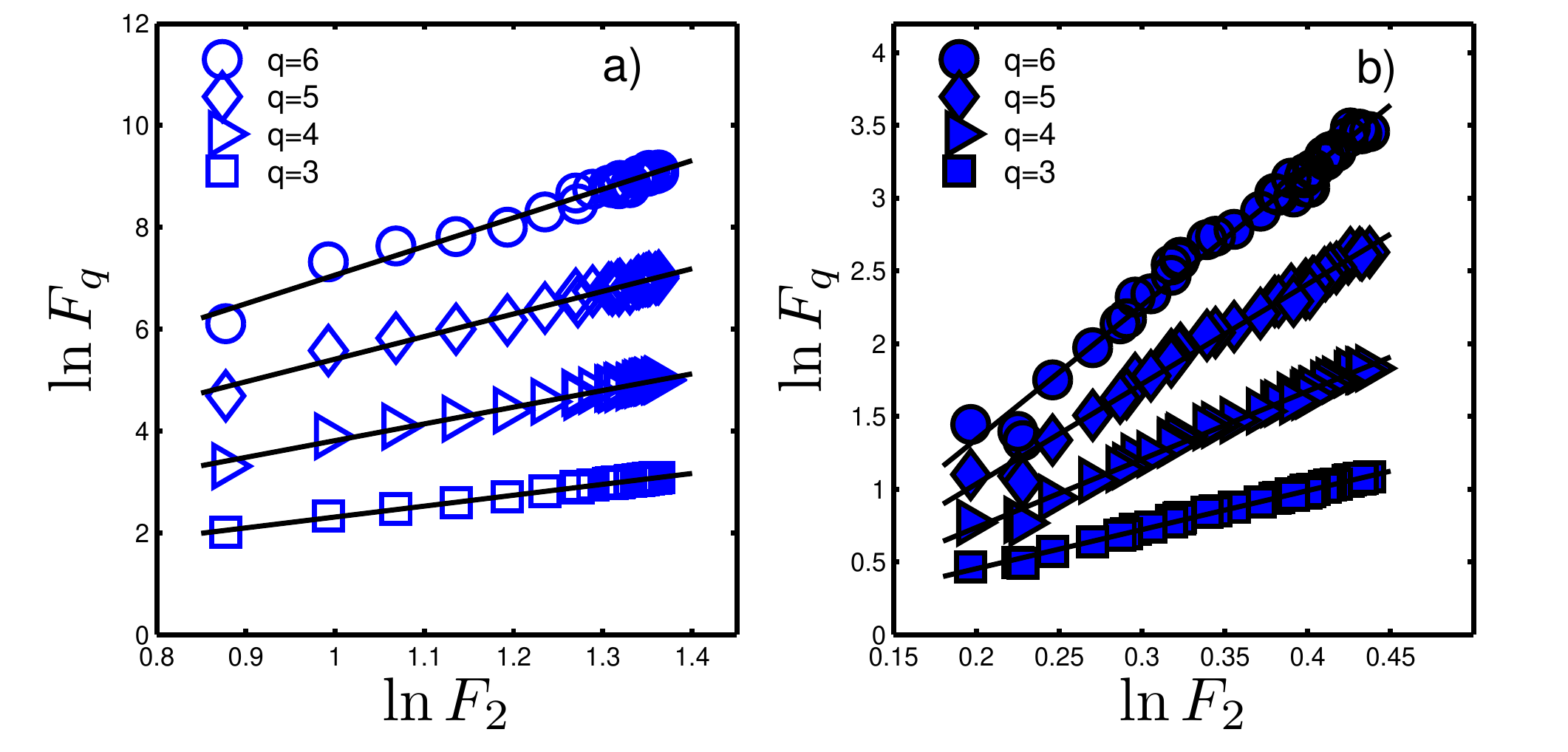}
     \caption{Log-log plot of $F_q$ as a function of $F_2$ for {\it Conference} data. (a) for contacts; and (b) for nodes.}
\end{figure}

It is known that $\gamma$ is approximately 1.3 for 2D Ising model undergoing a second-order phase transition \cite{HwaPRL}. The exponents $\gamma$ of contacts in {\it Prostitution} data $(\gamma = 1.341)$ and of nodes in {\it Conference} data $(\gamma = 1.345)$ are very close to this value.

\begin{figure}
    \centering
     \includegraphics[width=0.45\textwidth]{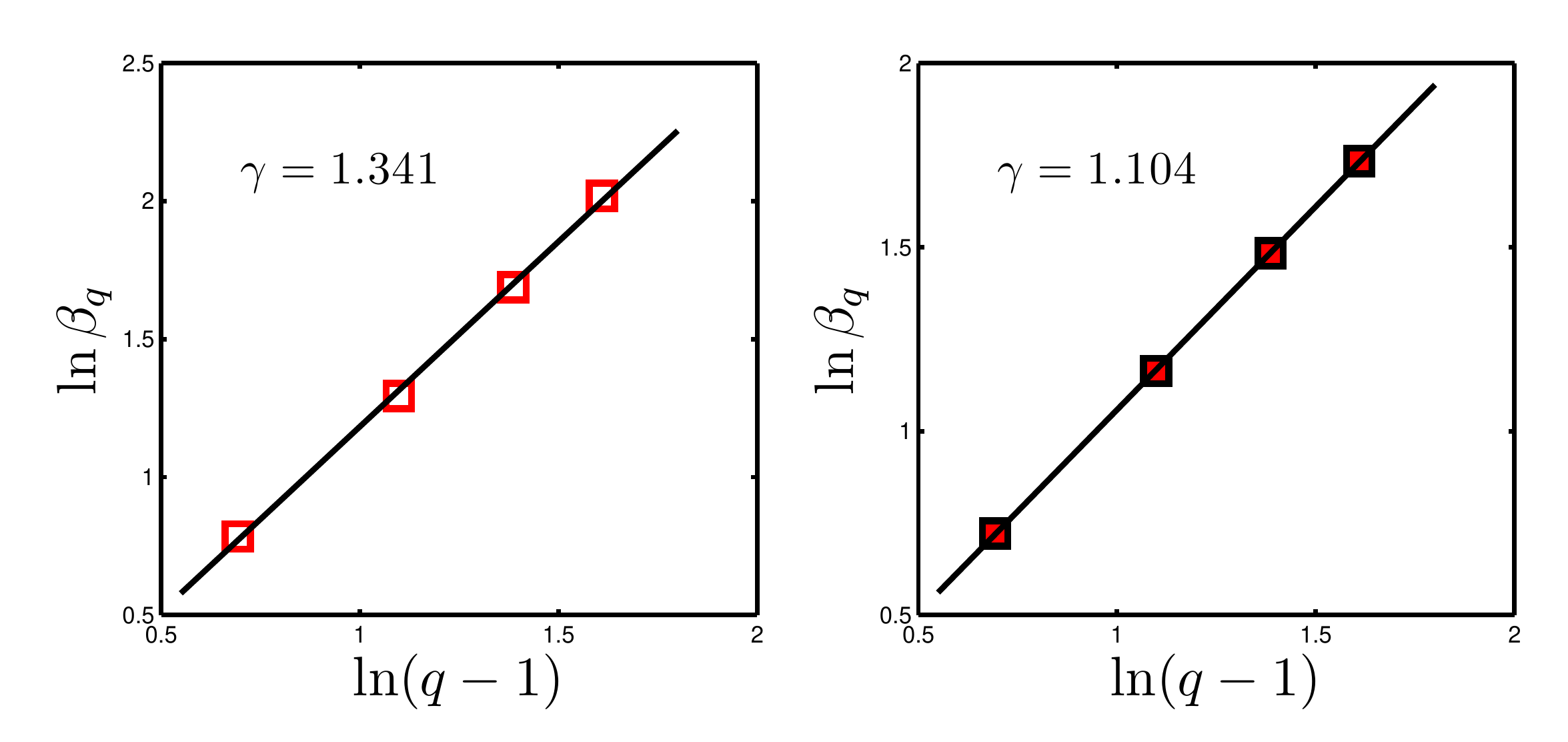}
     \caption{Scaling properties between $\beta_q$ and $(q-1)$ for {\it Prostitution} data. The scaling exponents are about: (a) 1.341 for contacts; and (b) 1.104 for nodes.}
\end{figure}

\begin{figure}
    \centering
     \includegraphics[width=0.45\textwidth]{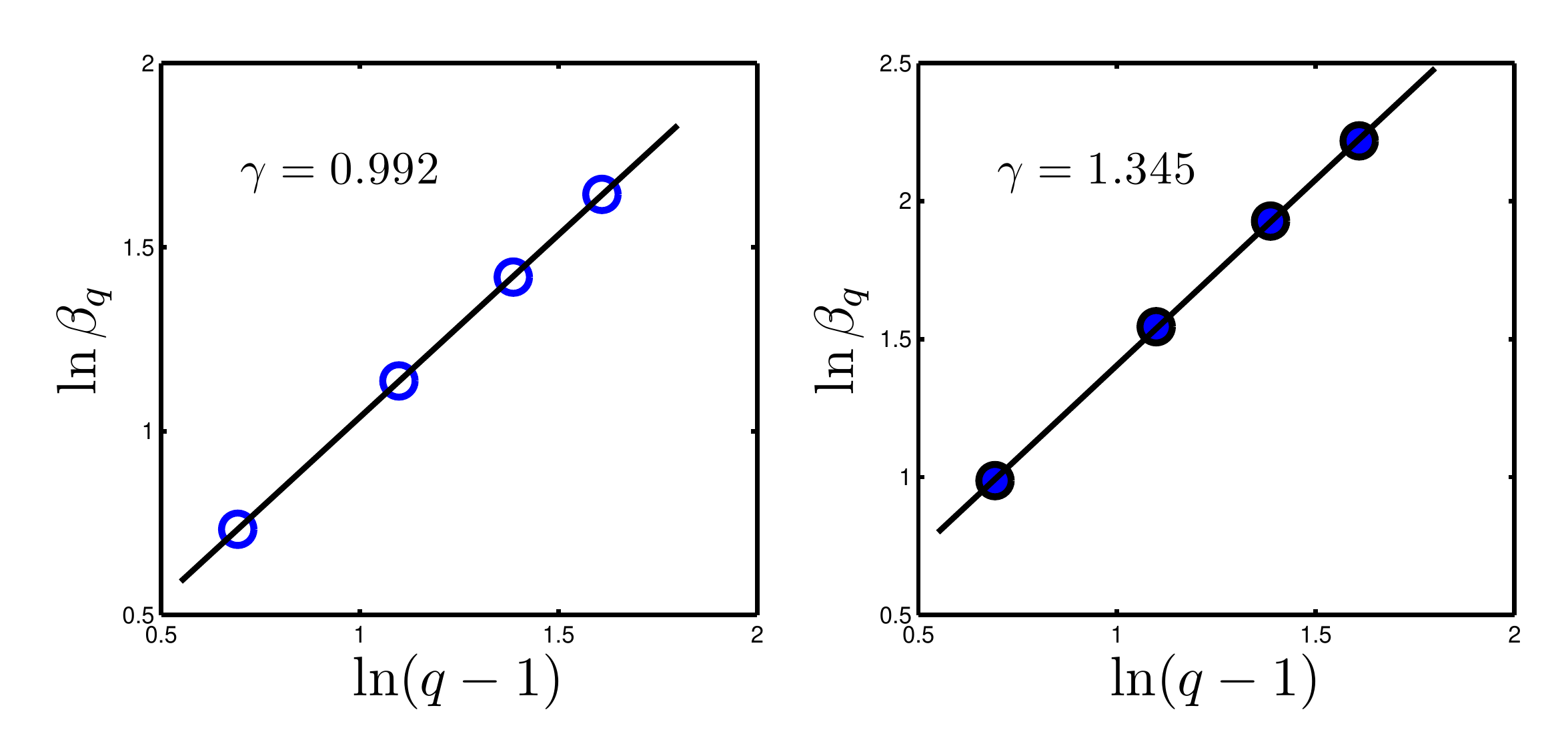}
     \caption{Scaling properties between $\beta_q$ and $(q-1)$ for {\it Conference} data. The scaling exponents are about: (a) 0.992 for contacts; and (b) 1.345 for nodes.}
\end{figure}

\section{Conclusions}

The factorial moments analyses are performed to study the scaling properties for fluctuations of contacts and nodes in temporal networks based on empirical data sets. The phenomena of intermittency $F_q \sim M^{\alpha_q}$ have been observed for all orders $q$ in the fluctuations of contacts and nodes for both {\it Prostitution} and {\it Conference} data sets. The result indicates that the system has self-similar structure and the fluctuations are not purely random, but have dynamical and correlated behaviors embedded in the system. A more general scaling relationship between $F_q$ and $F_2$ has also been presented, $F_q \sim F_2^{\beta_q}$. We further find that $\beta_q$ scales with $q$ as $\beta_q \sim (q-1)^{\gamma}$. The exponents $\gamma$ for nodes in the {\it Prostitution} data and for contacts in the {\it Conference} data are very close to that for 2D Ising model. The other exponents $\gamma$ are not.

Still, there are some issues to be addressed. First, what is the driving mechanism(s) behind these scaling properties of fluctuations in temporal networks? Second, why are some scaling exponents close to that of Ginzburg-Landau second-order phase transition? Are they belong to the same universal class? All these topics cannot be covered in this paper and will be discussed later.

The scaling invariances of fluctuations shed light on the temporal correlations of contact series and provide a new sight into understanding the influence of time dimension in temporal networks.

\section*{Acknowledgments}
We are grateful to Professor Petter Holme for providing the data. This work was supported in part by the programme of Introducing Talents of Discipline to Universities under No. B08033, and by the self-determined research funds of CCNU from the colleges' basic research and operation of MOE under No. CCNU2015A05046.

\bibliographystyle{plain}
\bibliography{bibliography}

\end{document}